# From UML Specification into Implementation Using Object Mapping

Rosziati Ibrahim

Faculty of Information Technology and Multimedia

Universiti Tun Hussein Onn Malaysia

Parit Raja, Batu Pahat, 86400, Johor, Malaysia

Tel: 60-7-453-8001     E-mail: rosziati@uthm.edu.my

*This research is supported by the Science Fund Grant Vote S014*

**Abstract**

In information systems, a system is analyzed using a modeling tool. Analysis is an important phase prior to implementation in order to obtain the correct requirements of the system. During the requirements phase, the software requirements specification (SRS) is used to specify the system requirements. Then, this requirements specification is used to implement the system. The requirements specification can be represented using either a structure approach or an object-oriented approach. A UML (Unified Modeling Language) specification is a well-known for representation of requirements specification in an object-oriented approach. In this paper, we present one case study and discuss how mapping from UML specification into implementation is done. The case study does not require advanced programming skills. However, it does require familiarity in creating and instantiating classes, object-oriented programming with inheritance, data structure, file processing and control loop. For the case study, UML specification is used in requirements phase and Borland C++ is used in implementation phase. Based on the case study, it shows that the proposed approach improved the understanding of mapping from UML specification into implementation.

## 1. Introduction

In information systems, modeling of any new systems can be done using a standard methodology. The model used can be either of structured approach or object-oriented approach. Structured approach uses diagrams such as entity relationship diagrams (ERD) and context diagrams to model and analyze the system requirements (Hoffer et al., 2008). Object-oriented approach, on the other hand, uses diagrams such as use-case diagrams and class diagrams to model and analyze the system requirements (Dennis et al., 2005). Unified Modeling Language (UML) is one of the modeling tools that are often used for object-oriented approach. UML assumes a process that is use-case driven, architecture-centered, iterative and incremental (Bahrami, 1999). UML is a standard language for visually describing the structure and behavior of a system. Therefore, during analysis, the system requirements are transformed into UML specification using diagrams. These diagrams include the Use-Case Diagram, the Class Diagram, the Interaction Diagram, the Communication Diagram, the Activity Diagram, the State Diagram, the Component Diagram and the Deployment Diagram (Miller, 2003). This paper illustrates a case study where students are thought to map requirements analysis phase into implementation phase by using the UML specification. From this case study, based on the results from student evaluation forms, the results show that students improved their understanding on object mapping from UML specification into implementation.

The rest of this paper is organized as follows. Section 2 and Section 3 briefly review the basic of object-oriented programming and the UML specification, respectively. Section 4 shows a case study on how to map from UML specification into implementation and Section 5 concludes the paper.

## 2. The Basics of Object-Oriented Programming

Object-oriented programming (OOP) is one of the ways in organizing and developing software. The foundation of OOP languages goes back to Simula and Smalltalk. In Simula, a program is a collection of objects. The notion of objects was first introduced in the Simula language designed in the late 60s (Micallef, 1988). However, OOP did not emerge as a new programming paradigm until Smalltalk came along in the late 70s. In object-oriented approach, everything that can be touched is considered as an object. For example, a person, a car, a doll and a house can be considered as objects. An object is an entity. It contains attributes and provides services. Figure 1 shows an example of an object.

Before an object can be used, it needs to be created (instantiated). An object is created from a class. A class is a template from which objects are instantiated. A class consists of fields and methods. Fields are data for class instances and methods are operations that access the data. Diagram 1 shows a class Car of an object MyCar from Figure 1.

When an object wants to communicate with another object, that object sends a message to another object. Objects communicate by sending messages. It does not allow violation of data within the object. Figure 2 shows how objects interact with each other by means of messaging.

In C++, for example, if an object wants another object to do some work on its behalf, it can send a message to that object and that object selects the appropriate method to invoke. Therefore, a message sent to an object must be corresponded to a method invocation of the object. A method invocation is similar to a method call in C++. Figure 3 shows an example of object communications in C++.

In reference to Figure 3, let us consider for example, class Student who wants to communicate with class Faculty. Recall that we have to instantiate class Student and class Faculty first before messages can be sent. Let us assume that we instantiated class Student to Angelina and class Faculty to FCSIT (Code for Faculty of Computer Science and Information Technology). Then, a message can be sent from Angelina object. Here method Register is called (denotes by Angelina.Register). From method Register, method RegisterCourse is indirectly called from class Faculty via FCSIT object (denotes by FCSIT.RegisterCourse). In object-oriented approach, this communication is also known as an association between objects. Figure 4 shows the association between these two objects.

In OOP, inheritance is widely used because it allows reusable of existing source codes. Inheritance is a relationship between classes where one class is parent class (superclass) of its child class (subclass). Inheritance is also used to communicate the concept that one class can inherit part of its behaviour and data from another class. For example, a subclass of a program can inherit some code from its superclass. Figure 5 shows an example of inheritance. Further details regarding object-oriented programming using inheritance can be found in (Rosziati et al., 2006).

### 3. Unified Modeling Language (UML) Specification

UML is a standard language for modeling of a system. UML is used to visually specify the structure and behavior of a system. The system requirements are captured and then converted into UML specifications which are represented by UML diagrams. A use case diagram is used to specify requirements of the system. In a use-case diagram, two important factors are used to describe the requirements of a system. They are actors and use cases. Actors are external entities that interact with the system and use cases are the behavior (or the functionalities) of a system (Bahrami, 1999). The use cases are used to define the requirements of a system.

The class diagram, on the other hand, is the main static analysis diagram (Bahrami, 1999). It shows the static structure of the model for the classes and their relationships. They are connected to each other as a graph. Each class has its own internal structures and its relationships with other classes.

### 4. Object Mapping from UML Specification into Implementation: A Case Study

Mapping from UML specification into implementation involves using UML diagrams and converting them into program source codes. In this paper, we present a case study and show how to come up with a proper requirements analysis before implementing the system. The problem statement is analyzed in order to get UML specifications. Then UML specifications are used for conversion into program source codes.

> *Problem Statement*: Write a program for Research Management System. This program would be able to read from a file that contains information regarding the researcher vote details (such as the name, vote number and balance, and his/her password). The program then allows a user to login into the system. If the user is able to login into the system successfully, then the user is able to commit making the order for the item that he/she wishes to buy through the system and/or check the balance from his/her research vote as well as display the information details of the user. If the user is an administrator, the user will be able to view the order file.

From the problem statement above, we can convert it into programming style (which is known as pseudo-code). If we understand the problem statement carefully, these are the tasks that we are supposed to do:

    Read from a file
    Login into the system

      If successful, then do various activities such as
- Commit Order
- Check Vote Balance
- Display information details
- View Order File

      End

Then, from the pseudo-code, we would be able to come up with the requirements of the system such as to login into the system, to commit making order for an item, to check vote balance, to view order file and to display information details of the user. From these requirements, we come up with a use-case diagram which has six use cases: *Login, Commit, CheckBalance, DisplayDetails, ViewOrder* and *RecordOrder*. Two actors can use this system. They are Researcher and Administrator. Researcher would be able to *Login*, *Commit*, *CheckBalance* and *DisplayDetails* and Administrator would be able to *Login* and *ViewOrder*. The *Commit* use case, on the other hand, extends the *RecordOrder* use case. Diagram 2 shows the use-case diagram of our case study.

From the use-case diagram, we can proceed creating classes for a class diagram. Creating useful and meaningful classes are important in order to implement the correct classes. Using the use-case diagram from Diagram 2, we can create four important classes. The classes are Research, Activity, Order and System. From these classes we can then create relationships between classes and develop a class diagram. Diagram 3 shows class diagram for our case study.

From Diagram 3, it shows the relationships between classes where class System acts like a system that initiates the whole program by having three functions namely *ReadFile*, *Login* and *Menu*. Then, class Activity will do the activities from the Researcher and Administrator based from the activity selected from function *Menu* from class System (that is from function *Menu*, either function *Commit*, *CheckBalance*, *DisplayDetails* or *ViewOrder* can be invoked). Recall that from the use-case diagram in Diagram 2, Researcher has *Login, Commit, CheckBalance* and *DisplayDetails* as use cases, Administrator has *Login* and *ViewOrder* as use cases and the *Commit* use case extends the *RecordOrder* use case. These use cases are transformed into functions in class diagram in Diagram 3. For class Activity, we can have functions such as *Commit*, *CheckBalance* and *DisplayDetails*. However, these functions depend on attributes from class Research. Therefore, we can also use inheritance mechanism between classes Activity and Research where class Activity is a subclass of class Research. Since class Activity inherits class Research, class Activity can reuse data declared in class Research for the purpose of setting the data in class Order. This simplifies the data used and the program written as well.

Diagram 3 also shows class Activity, which inherits from class Research. We also add two more functions, namely *GetActivityType* and *GetAnotherAct*. Function *GetActivityType* is used to get the type of activities from a user and function *GetAnotherAct* is used to obtain an input from a user as to whether he or she wants to continue using the system (program).

Note that we have a class Order which is used from class Activity. This implies that class Activity requests to communicate with class Order via message. That is, class Activity will send a message by means of a method invocation to class Order. This communication is also known as association between two classes.

We can now implement our programs according to class diagram shown in Diagram 3. The mapping from UML specification into implementation is followed strictly. All the program source codes in C++ adhere to classes declared in Diagram 3. Note that, only the header files are shown in this paper.

Diagram 4 shows how class System is implemented. Class System has one attribute (Data is declared using structure in C++ for details of the user) and three methods. Class System acts like a task that the system supposes to do (more like the pseudo-code of the system). Therefore, for class System, we can have functions such as *ReadFile*, *Login* and *Menu*. Once the system finishes reading a file and login is successful, the system will proceed with executing one of the activities from function Menu (see Rosziati (2008) for full program source codes).

Diagram 5 shows how classes Research and Activity are implemented using inheritance according to Diagram 3. From Diagram 5, class Activity has no attributes. However, because of inheritance, class Activity can use all attributes from class Research, which is its superclass. Note that from Diagram 3, because of association between classes System and Activity, functions from class Activity will be invoking from the activity selected from function Menu in class System.

Class Order records all the commit orders from all the researchers. Therefore, for class Order, we can have functions such as *RecordOrder* and *ViewOrder*. Diagram 6 shows the class Order. Note that further details regarding the full program source codes can be found in (Rosziati, 2008).

Recall that from use-case diagram in Diagram 2, use case Commit extends use case RecordOrder. Therefore, from Diagram 6, function *RecordOrder* will be called via function *Commit* from class Activity.

The object mapping from UML specification into implementation shows the importance of getting the correct requirements of a system. We present this case study to our students and show them how to implement the system from requirements analysis using UML specifications (in particular using use-case diagram and class diagram). Students appreciated the hardcore programming and this case study has made students appreciated the rigorous analysis of the system that needs to be implemented. Of course, most industrial programming projects are more complex than this case study. However, this case study emphasizes on students programming skills as well as analytical thinking on requirements analysis. At the end of the semester, from students evaluation forms, we also received testimony from our students of the benefits of completing this case study (39 out of 45 students (87%) agreed that the case study improved their understanding of object mapping from UML specification into implementation). They also suggested that more such case studies be included in future classes. This case study motivates students to work from initial phase (getting correct requirements of the system) until the implementation phase.

## 5. Conclusion

This paper has discussed the requirements analysis phase, the implementation phase and the object mapping from UML specification into implementation using object-oriented approach. It is crucial to show to students the importance of the object mapping in order to get consistency between requirements analysis and implementation. In other cases, requirements analysis does not go along with implementation. Here, software developers have to change the analysis due to changing of the program source codes during implementation. These will create errors such as system does not meet the system requirements. If analysis is strictly adhered too, changes can always be avoided and implementation would be easier. This case study shows how object mapping is done and how changes between two phases are kept to minimum.

The case study for Research Management System also shows that before implementation, a rigorous analysis is important in order to capture system requirements. What is lacking in ordinary system implementation, people tend to forget about the importance of requirements analysis phase. This case study shows to students the important of requirements analysis phase. It also shows the object-oriented approach of implementing the system using the capability of inheritance in object-oriented programming.

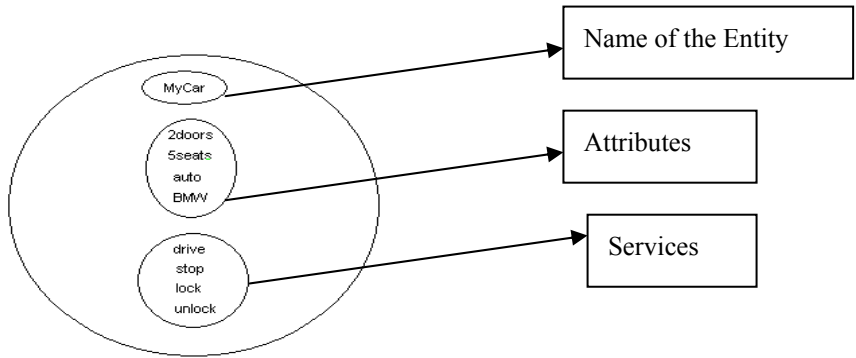

Figure 1. Example of an Object - MyCar

Diagram 1. Class Car

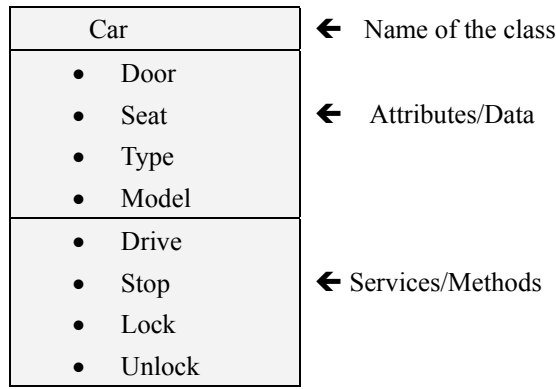

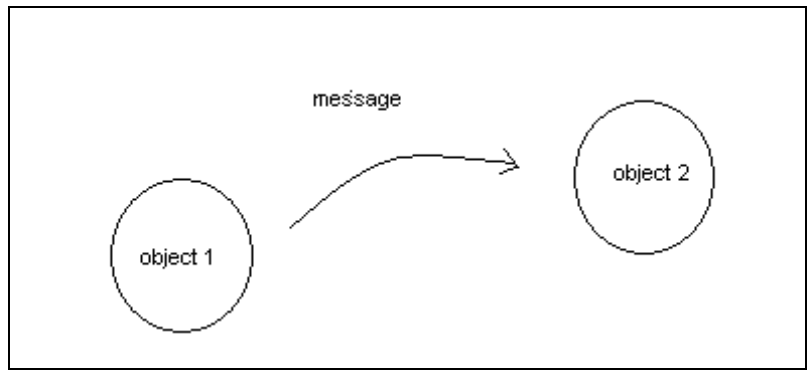

Figure 2. Objects Communications

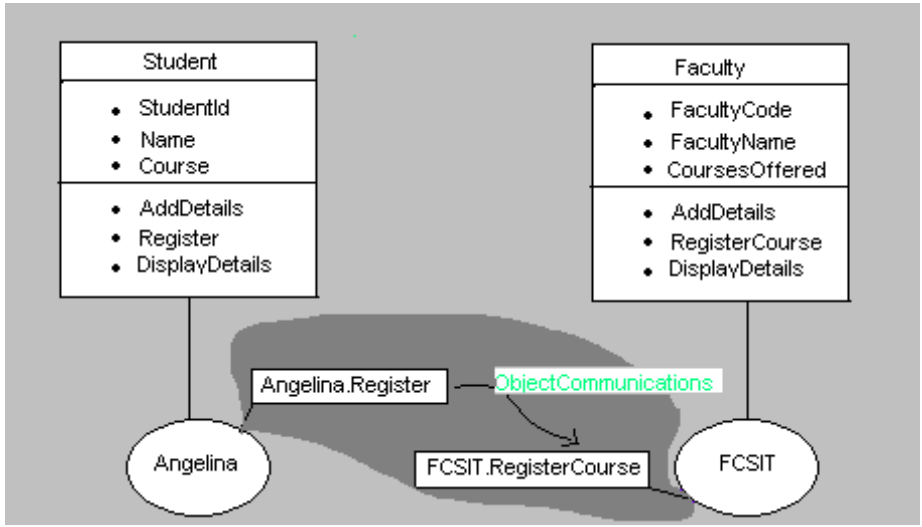

Figure 3. Example of Object Communication

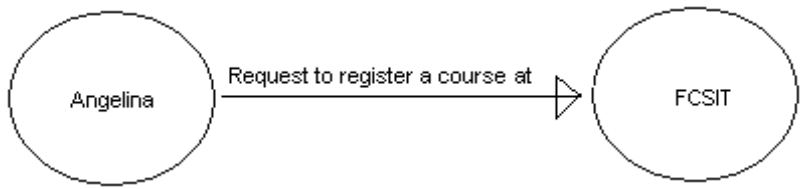

Figure 4. The association between two objects

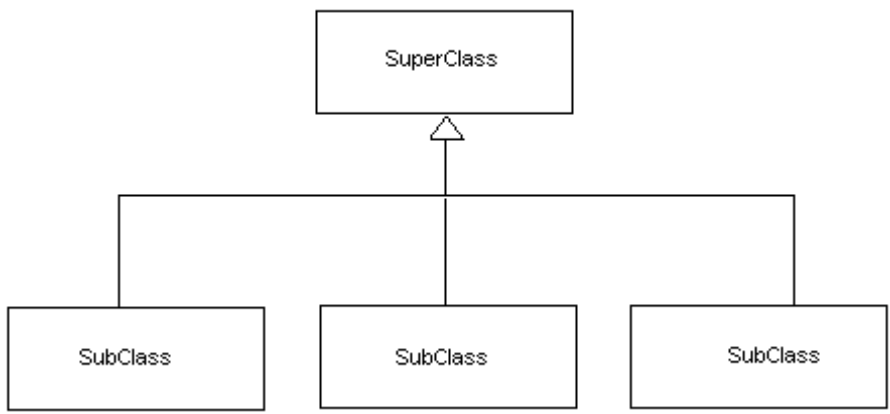

Figure 5. An Example of Inheritance

Diagram 2. Use-Case Diagram for Research Management System

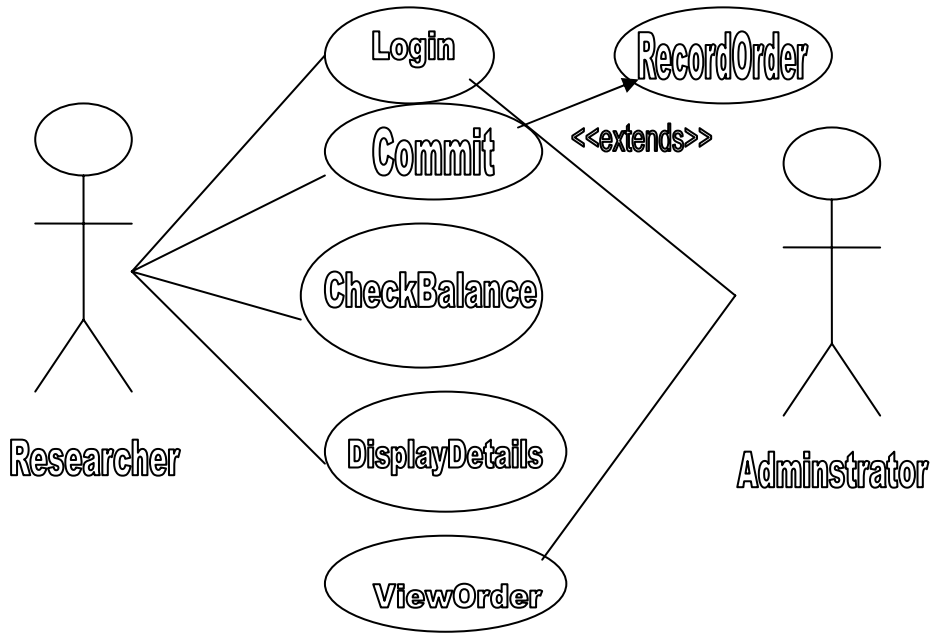

Diagram 3. Relationships of Classes (Class Diagram)

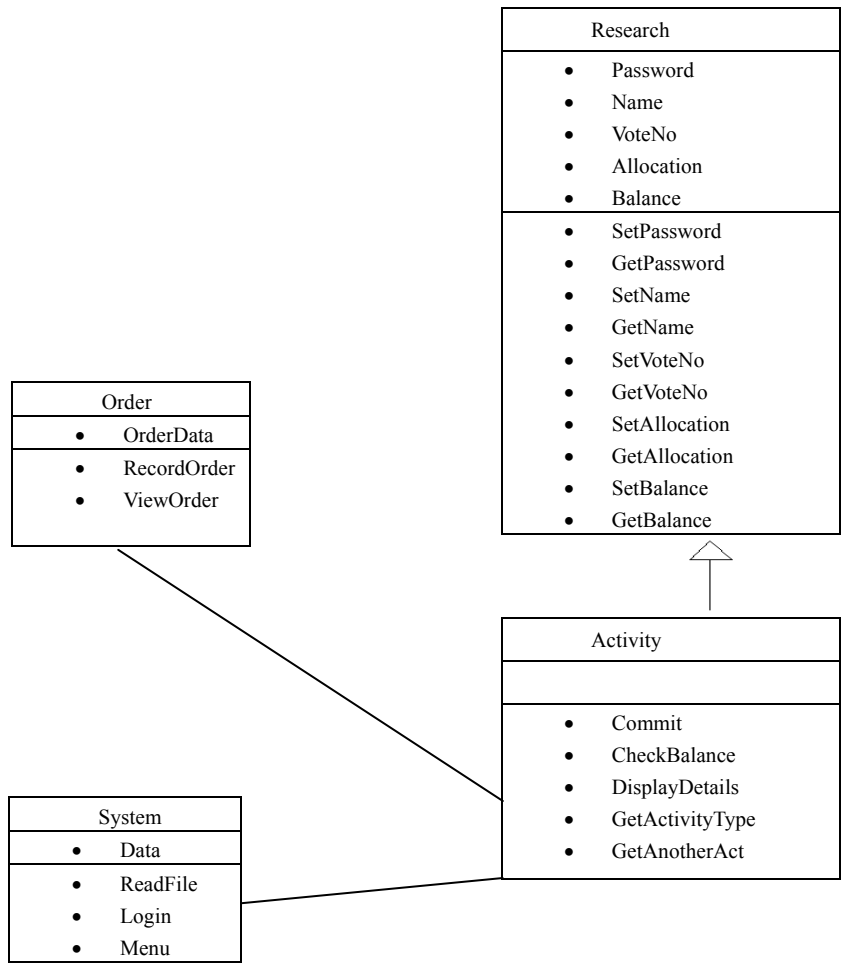

Diagram 4. Mapping of Class System into Implementation

| UML Specification | Implementation |
|---|---|
| System | **Class** System { |
| • Data | **private**:<br>    **struct** Data {    //Data for information details for user<br>        **char** password[7];<br>        **char** name[20];<br>        **char** votenum[8];<br>        **int** allocation;<br>        **int** balance;<br>    } dat; |
| • ReadFile<br>• Login<br>• Menu | **public**:<br>    **void** ReadFile();<br>    **void** Login();<br>    **void** Menu();<br>}; // class System |

Diagram 5. Mapping of Class Activity using inheritance with Implementation

| Analysis | Implementation |
|---|---|
| Research | **class** Research { |
| • Password<br>• Name<br>• VoteNo<br>• Allocation<br>• Balance | **Private**:<br>    **char** Password[7];<br>    **char** Name[20];<br>    **char** VoteNo[8];<br>    **int** Allocation;<br>**protected**:<br>    **int** Balance; |
| • SetPassword<br>• GetPassword<br>• SetName<br>• GetName<br>• SetVoteNo<br>• GetVoteNo<br>• SetAllocation<br>• GetAllocation<br>• SetBalance<br>• GetBalance | **public**:<br>    Research ();   //constructor<br>    **void** SetPassword(**char** ResPassword[7]);<br>    **char** *GetPassword();<br>    **void** SetName(**char** ResName[20]);<br>    **char** *GetName();<br>    **void** SetVoteNo(**char** ResNum[8]);<br>    **char** *GetVoteNo();<br>    **void** SetAllocation(**int** Aloc);<br>    **int** GetAllocation ();<br>    **void** SetBalance(**int** bal);<br>    **int** GetBalance ();<br>}; //class Research |
| Activity | **class** Activity : **public** Research {        //inheritance |
|  |  |
| • Commit<br>• CheckBalance<br>• DisplayDetails<br>• GetActivityType<br>• GetAnotherAct | **public**:<br>    Activity();   //constructor<br>    **void**    Commit(**int** Amt);<br>    **void**    CheckBalance();<br>    **void**    DisplayDetails();<br>    **int**     GetActivityType();<br>    **char**    GetAnotherAct();<br>}; //class Activity |

Diagram 6. Mapping of Class Order into Implementation

| *Analysis* | *Implementation* |
|---|---|
| Order | **class** Order { |
| • OrderData | **private**:<br>   **struct** OrderData {<br>  **char** Name[25];<br>  **char** VoteNo[8];<br>  **char** OrderDetail[25];<br>  **int** Amount;<br>  } stdata; |
| • RecordOrder<br>• ViewOrder | **public**:<br>  **void** RecordOrder();<br>  **void** ViewOrder();<br>};   // class Order |